\pdfoutput=1
\documentclass[12pt]{article}
\usepackage{graphicx}
\usepackage{amsmath, amssymb}

\newcommand{\mbf}[1]{\mathbf{#1}}

\renewcommand{\bar}[1]{\overline{#1}}

\def\Dslash{\raise.15ex\hbox{/}\kern-.7em D}
\def\Pslash{\raise.15ex\hbox{/}\kern-.7em P}

\begin {document}
\begin{flushright}
{\small
SLAC--PUB--13428\\
October 2008}
\end{flushright}

\vspace{10pt}

\centerline{\LARGE \bf The AdS/CFT Correspondence}

\vspace{7pt} 

\centerline{\LARGE \bf and Light-Front QCD}

\vspace{20pt}

\centerline{\bf {
Stanley J. Brodsky$^{a}$
\!\footnote{Speaker}
 and
Guy F. de T\'eramond$^{b}$}
}
\vspace{10pt}

{\centerline {$^{a}${Stanford Linear Accelerator Center, 
Stanford University, Stanford, CA 94309, USA}}

\vspace{3pt}

{\centerline {$^{b}${Universidad de Costa Rica, San Jos\'e, Costa Rica}}

\vspace{15pt}

\begin{abstract}

We identify an invariant  light-front coordinate $\zeta$ which
allows the separation of the dynamics of quark and gluon binding
from the kinematics of constituent spin and internal orbital angular
momentum. The result is a single-variable light-front Schr\"odinger
equation for QCD which determines the eigenspectrum and the
light-front wavefunctions of hadrons for general spin and orbital
angular momentum. This frame-independent light-front wave equation is equivalent to
the equations of motion which describe the propagation of spin-$J$
modes on  anti-de Sitter (AdS) space.  Light-front holography is a remarkable feature of AdS/CFT:  it allows hadronic amplitudes in the AdS fifth dimension to be mapped to frame-independent light-front wavefunctions of hadrons in physical space-time, thus providing a  relativistic description of hadrons at the amplitude level.   In principle, the model can be systematically improved by diagonalizing the full QCD light-front Hamiltonian on the AdS/QCD basis.
Quark and gluon hadronization can be computed at the amplitude level by convoluting the off-shell $T$ matrix calculated from the QCD light-front Hamiltonian with the hadronic light-front wavefunctions.
We also note the distinction between static observables such as the probability distributions  computed from the square of the light-front wavefunctions versus  dynamical observables such as the structure functions and the leading-twist single-spin asymmetries  measured in deep inelastic scattering which include the effects of initial and final-state interactions.

\end{abstract}

\begin{center}
{\it \small Presented at  LIGHT CONE 2008: Relativistic Nuclear and Particle Physics\\
      Mulhouse, France, July 7-11  2008
 }
\end{center}

\vfill

\newpage

\section{Introduction}

One of the most important theoretical tools in atomic physics is the
Schr\"odinger equation, which describes the quantum-mechanical
structure of atomic systems at the amplitude level. Light-front
wavefunctions (LFWFs) play a similar role in quantum chromodynamics
(QCD), providing a fundamental description of the structure and
internal dynamics of hadrons in terms of their constituent quarks
and gluons. The natural concept of a wavefunction for relativistic quantum field theories such as QCD is the light-front wavefunction $\psi_n(x_i, \mbf{k}_{\perp i}, \lambda_i)$ which specifies the  $n$ quark and gluon constituents of a hadron's Fock state as a function of the light-cone fractions $x_i = k^+/P^+ = (k^0+k^3)/(P^0+P^3)$ transverse momenta $\mbf{k}_{\perp i}$ and spin projections $\lambda_i$. The light-front wavefunctions of bound states in QCD are
relativistic generalizations of the Schr\"odinger wavefunctions of
atomic physics, but they are determined at fixed light-cone time
$\tau  = t +z/c$ -- the ``front form'' introduced by
Dirac~\cite{Dirac:1949cp} -- rather than at fixed ordinary time $t$.

When a flash from a camera illuminates a scene, each object is illuminated along the light-front of the flash; i.e., at a given $\tau$.   In contrast, setting the initial condition using conventional instant time $t$ requires simultaneous scattering of photons on each constituent. 
Thus it is natural to set boundary conditions at fixed $\tau$ and then evolve the system using the light-front Hamiltonian 
$P^- = P^0-P^3 = i {d/d \tau}.$  The invariant Hamiltonian $H_{LF} = P^+ P^- - \mbf{P}^2_\perp$ then has eigenvalues $\mathcal{M}^2$ where $\mathcal{M}$ is the physical mass.   Its eigenfunctions are the light-front (LF) eigenstates whose Fock state projections define the light-front wavefunctions.   

A remarkable feature of LFWFs is the fact that they are frame
independent; i.e., the form of the LFWF is independent of the
hadron's total momentum $P^+ = P^0 + P^3$ and $\mbf{P}_\perp.$  The light-front formalism for gauge theories in
light-cone gauge  $A^+ = 0$  is particularly useful in that there are no ghosts  and the gluon
polarization is purely transverse: $S^z_g = \pm 1$.
Thus one has a direct physical interpretation of  orbital angular
momentum.  
The constituent spin and orbital angular momentum properties of the
hadrons are also encoded in the LFWFs. For example, the internal
spin and orbital angular momentum is conserved for each $n$-particle LF Fock
state:
$\sum^n_{i=1}S^z_i + \sum^{n-1}_{i=1}L^z_i = J^z,$
since there are $n-1$ relative orbital angular momentum. Since the plus momenta are conserved and positive, the vacuum in front form is trivial except for $k^+$ zero modes. For example, in the case of the Higgs theory, a $c$-number LF  zero mode 
constant~\cite{Srivastava:2002mw} replaces the vacuum condensate of the instant form.   The simple structure of the light-front vacuum allows an unambiguous definition of the partonic content of a hadron.

Light-front wavefunctions are the fundamental process-independent amplitudes which encode hadron properties  in terms of their quark and gluon degrees of
freedom, predicting dynamical quantities such as spin correlations, form factors, structure functions, generalized parton distributions, and exclusive scattering amplitudes.  Meson and baryon light-front wavefunctions can
be measured in diffractive di-jet and tri-jet reactions,
respectively. One of the most important advantages of the light-front formalism is
that spacelike form factors can be represented as simple overlap
integrals of the LF Fock state wavefunctions $\psi_n$ and
$\psi_{n^\prime}$ with $n^\prime =n$; i.e.,  the Drell-Yan-West
formula. This is in dramatic contrast to the usual instant form
result which requires the inclusion of contributions where the
current couples to vacuum processes.  
Thus knowing the
wavefunction of a hadron at fixed time $t$ is not sufficient to
determine the form factors and other properties of the hadron. In
addition, one must also be able to compute the boosted instant form wavefunction,
which requires solving a complex dynamical problem. In fact, boosted
wavefunctions are only known at weak coupling and even then are more
complicated than the product of Melosh or Wigner transformations of
the individual constituent spinors. In contrast, the light-front wavefunctions  of a
hadron are independent  of the 
momentum of the hadron, and they are thus boost invariant. 
The generalized parton distributions measured in deep inelastic
Compton scattering $\gamma^*(q) p \to \gamma(k) p^\prime$ in the handbag approximation can be
written as the overlap of light-front wavefunctions~\cite{Brodsky:2000xy}.  

\section{ A Single-Variable Light-Front Schr\"odinger Equation for QCD~\cite{deTeramond:2008ht}}

A key step in the analysis of an atomic system such as positronium
is the introduction of the spherical coordinates $r, \theta, \phi$
which  separates the dynamics of Coulomb binding from the
kinematical effects of the quantized orbital angular momentum $L$.
The essential dynamics of the atom is specified by the radial
Schr\"odinger equation whose eigensolutions $\psi_{n,L}(r)$
determine the bound-state wavefunction and eigenspectrum. Here
we show that there is an analogous invariant light-front
coordinate $\zeta$ which allows one to separate the essential
dynamics of quark and gluon binding from the kinematical physics of
constituent spin and internal orbital angular momentum. The result
is a single-variable light-front Schr\"odinger equation for QCD
which determines the eigenspectrum and the light-front wavefunctions
of hadrons for general spin and orbital angular momentum.  Conversely, 
this analysis can be applied to atomic physics, providing an elegant formalism for relativistic atoms.

The connection between light-front QCD and the description of
hadronic modes on AdS space is physically compelling and
phenomenologically successful. To  a first approximation
light-front QCD is formally equivalent to an effective gravity
theory on AdS$_5$. To prove this, we show that  the LF Hamiltonian
equations of motion of QCD lead to an effective LF wave equation for
physical modes  $\phi(\zeta)$ which encode the hadronic properties.
This LF wave equations carry the orbital angular momentum quantum
numbers and are equivalent to the equations of motion which describe
the propagation of spin-$J$ modes on AdS space. This allows us to
formally establish  a gauge/gravity correspondence between an
effective gravity theory defined on AdS$_5$ and light front QCD at
its asymptotic boundary.

To simplify the discussion we will consider a two-parton hadronic
bound state. In the case of massless constituents the LF Hamiltonian equation of motion of QCD leads to the equation
\begin{eqnarray} \nonumber
\mathcal{M}^2  &\!\!=\!\!&  \int_0^1 \! d x \! \int \!  \frac{d^2
\mbf{k}_\perp}{16 \pi^3}   \,
  \frac{\mbf{k}_\perp^2}{x(1-x)}
 \left\vert \psi (x, \mbf{k}_\perp) \right \vert^2  + {\rm interactions} \\  \label{eq:Mb}
  &\!\!=\!\!& \int_0^1 \! \frac{d x}{x(1-x)} \int  \! d^2 \mbf{b}_\perp  \,
  \psi^*(x, \mbf{b}_\perp)
  \left( - \mbf{\nabla}_{ {\mbf{b}}_{\perp \ell}}^2\right)
  \psi(x, \mbf{b}_\perp)   +  {\rm interactions}.
 \end{eqnarray}
The functional dependence  for a given Fock state is
given in terms of the invariant mass
$
 \mathcal{M}_n^2  = \Big( \sum_{a=1}^n k_a^\mu\Big)^2 = \sum_a \frac{\mbf{k}_{\perp a}^2}{x_a}
 \to \frac{\mbf{k}_\perp^2}{x(1-x)} \,,
$
 the measure of the off-mass shell energy~ $\mathcal{M}^2 - \mathcal{M}_n^2$.
 Similarly in impact space the relevant variable for a two-parton state is  $\zeta^2= x(1-x)\mbf{b}_\perp^2$.
Thus, to first approximation  LF dynamics  depend only on the boost
invariant variable $\mathcal{M}_n$ or $\zeta$ and hadronic
properties are encoded in the hadronic mode $\phi(\zeta)$:
 $ \psi(x, \mbf{k}_\perp) \to \phi(\zeta)$.
 We choose the normalization of  the LF mode $\phi(\zeta) = \langle \zeta \vert \phi \rangle$ with
$
 \langle\phi\vert\phi\rangle = \int \! d \zeta \,
 \vert \langle \zeta \vert \phi\rangle\vert^2 = 1.
$
 Comparing with the LFWF normalization, we find the functional relation:  
 $\frac{\vert \phi \vert^2}{\zeta} = \frac{2 \pi}{x(1-x)} \vert\psi(x, \mbf{b}_\perp)\vert^2$, which is the same result
 found in~\cite{Brodsky:2006uqa, Brodsky:2007hb} from the mapping of transition matrix elements
 for arbitrary values of the momentum transfer.

We can write the Laplacian operator in circular
cylindrical coordinates $\zeta = (\vec \zeta, \varphi)$ with $\vec
\zeta = \sqrt{x(1-x)} \mbf{b}_\perp$: $\nabla^2 = \frac{1}{\zeta}
\frac{d}{d\zeta} \left( \zeta \frac{d}{d\zeta} \right) +
\frac{1}{\zeta^2} \frac{\partial^2}{\partial \varphi^2}$, and factor
out the angular dependence of the modes in terms of the $SO(2)$
Casimir representation $L^2$ of orbital angular momentum in the
transverse plane: $\phi(\vec \zeta, \varphi) \sim e^{\pm i L
\varphi} \phi(\zeta)$. We find
\begin{eqnarray} \nonumber
\mathcal{M}^2  &\!\!=\!\!& \int \! d\zeta \, \phi^*(\zeta)
\sqrt{\zeta} \left( -\frac{d^2}{d\zeta^2} -\frac{1}{\zeta}
\frac{d}{d\zeta} + \frac{L^2}{\zeta^2}\right)
\frac{\phi(\zeta)}{\sqrt{\zeta}}  
 + \int \! d\zeta \, \phi^*(\zeta) U(\zeta) \phi(\zeta) \\ 
&\!\!=\!\!& \int \! d\zeta \, \phi^*(\zeta) \left(
-\frac{d^2}{d\zeta^2} - \frac{1 - 4L^2}{4\zeta^2} +U(\zeta)\right)
\phi(\zeta),
\end{eqnarray}
where all the complexity of the interaction terms in the QCD
Lagrangian is summed up in the effective potential $U(\zeta)$. The
light-front eigenvalue equation $H_{LF} \vert \phi \rangle =
\mathcal{M}^2 \vert \phi \rangle$ is thus a light-front wave
equation for $\phi$
\begin{equation} \label{eq:QCDLFWE}
\left(-\frac{d^2}{d\zeta^2} - \frac{1 - 4L^2}{4\zeta^2} + U(\zeta)
\right) \phi(\zeta) = \mathcal{M}^2 \phi(\zeta),
\end{equation}
an effective single-variable light-front Schr\"odinger equation
which is relativistic, covariant and analytically tractable. One can readily generalize the equations to allow for
the kinetic energy of massive quarks~\cite{Brodsky:2008pg}.

As the simplest example we consider a bag-like
model~\cite{Chodos:1974je} where the partons are free inside the
hadron and the interaction terms will effectively build confinement.
The effective potential is a hard wall: $U(\zeta) = 0$ if  $\zeta
\le \frac{1}{\Lambda_{\rm QCD}}$ and
 $U(\zeta) = \infty$ if $\zeta > \frac{1}{\Lambda_{\rm QCD}}$.
 However, unlike the standard bag model~\cite{Chodos:1974je}, boundary conditions are imposed on the
 boost-invariant variable $\zeta$, not on the bag radius at fixed time.
 If $L^2 \ge 0$ the LF Hamiltonian is positive definite
 $\langle \phi \vert H_{LF} \vert \phi \rangle \ge 0$ and thus $\mathcal M^2 \ge 0$.
 If $L^2 < 0$ the LF Hamiltonian is unbounded from below and the particle
 ``falls towards the center''. The critical value corresponds to $L=0$.
  The mode spectrum  follows from the boundary conditions
 $\phi \! \left(\zeta = 1/\Lambda_{\rm QCD}\right) = 0$, and is given in
 terms of the roots of Bessel functions: $\mathcal{M}_{L,k}^2 = \beta_{L, k} \Lambda_{\rm QCD}$.
 Since in the conformal limit $U(\zeta) \to 0$, Eq. (\ref{eq:QCDLFWE}) is equivalent to an AdS
 wave equation, the hard-wall LF model discussed here is equivalent to the hard wall model of
 Ref.~\cite{Polchinski:2001tt}. Likewise a two-dimensional transverse oscillator with
 effective potential $U(\zeta) \sim \zeta^2$ is equivalent to the soft-wall model of
 Ref.~\cite{Karch:2006pv} which reproduce the usual linear Regge trajectories.
 
 \section{Light-Front Holography}

Our analysis follows from recent developments in light-front
QCD~\cite{deTeramond:2008ht,Brodsky:2006uqa,Brodsky:2007hb,Brodsky:2008pg}
which have been inspired by the
AdS/CFT correspondence~\cite{Maldacena:1997re} between string states
in anti-de Sitter (AdS) space and conformal field theories (CFT) in
physical space-time. The application of AdS space and conformal
methods to QCD can be motivated from the empirical
evidence~\cite{Deur:2008rf} and theoretical
arguments~\cite{Brodsky:2008be} that the QCD coupling $\alpha_s(Q^2)
$ has an infrared fixed point at low $Q^2.$ The AdS/CFT
correspondence has led to insights into the confining dynamics of
QCD and the analytic form of hadronic light-front wavefunctions. As
we have shown recently, there is a remarkable mapping between the
description of hadronic modes in AdS space and the Hamiltonian
formulation of QCD in physical space-time quantized on the
light-front. This procedure allows string modes $\Phi(z)$ in the AdS
holographic variable $z$ to be precisely mapped to the light-front
wave functions  of hadrons in physical space-time in terms of a
specific light-front variable $\zeta$ which measures the separation
of the quark and gluonic constituents within the hadron (see fig.~\ref{figholog}). The
coordinate $\zeta$ also specifies the light-front  kinetic
energy and invariant mass of constituents. This mapping was
originally obtained by matching the expression for electromagnetic
current matrix elements in AdS space with the corresponding
expression for the current matrix element using light-front theory
in physical space time~\cite{Brodsky:2006uqa,Brodsky:2007hb}. More recently we have
shown that one obtains the identical holographic mapping using the
matrix elements of the energy-momentum tensor~\cite{Brodsky:2008pf},
thus providing an important consistency test and verification of
holographic mapping from AdS to physical observables defined on the
light front.
The resulting wavefunction [see fig.~\ref{figff} (a)]
displays confinement at large interquark
separation and conformal symmetry at short distances, reproducing dimensional counting rules for hard exclusive amplitudes.
The predictions for the spacelike pion form factor for the hard-wall and soft-wall models is shown in fig. \ref{figff} (b).

\begin{figure}[!]
\begin{center}
 \includegraphics[width=10cm]{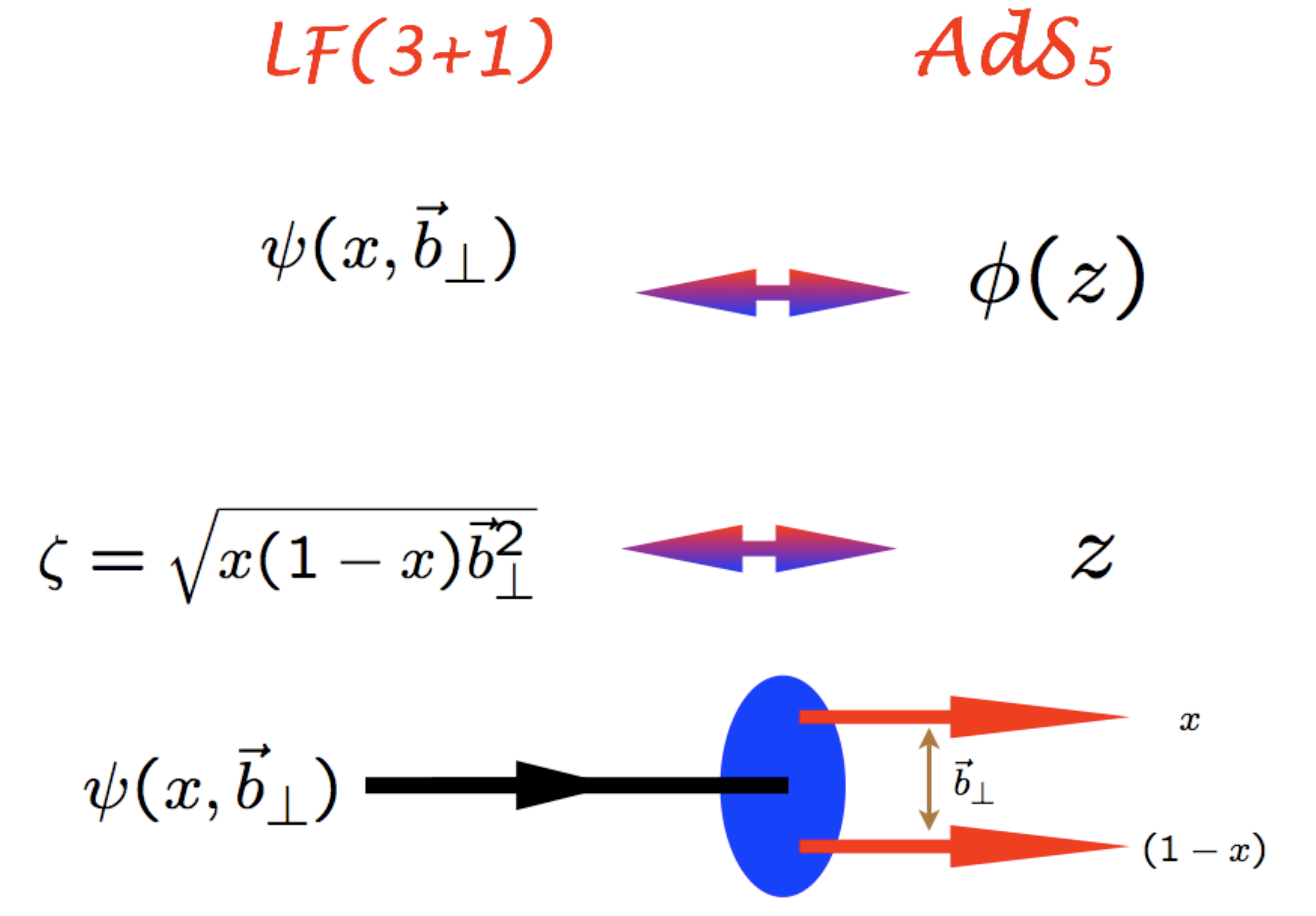}
 \end{center}
 \vspace{-10pt}
 \caption{Light-front holography for meson wavefunctions:  
 $\psi(x, \mbf{b}_\perp) = \sqrt{\frac{x(1-x)}{2 \pi \zeta}} \phi(\zeta)$. This  mapping is derived from the equality of the LF and AdS  formulae for current matrix elements.}
\label{figholog}  
\end{figure} 

\begin{figure}[!]
\begin{center}
\includegraphics[width=12cm]{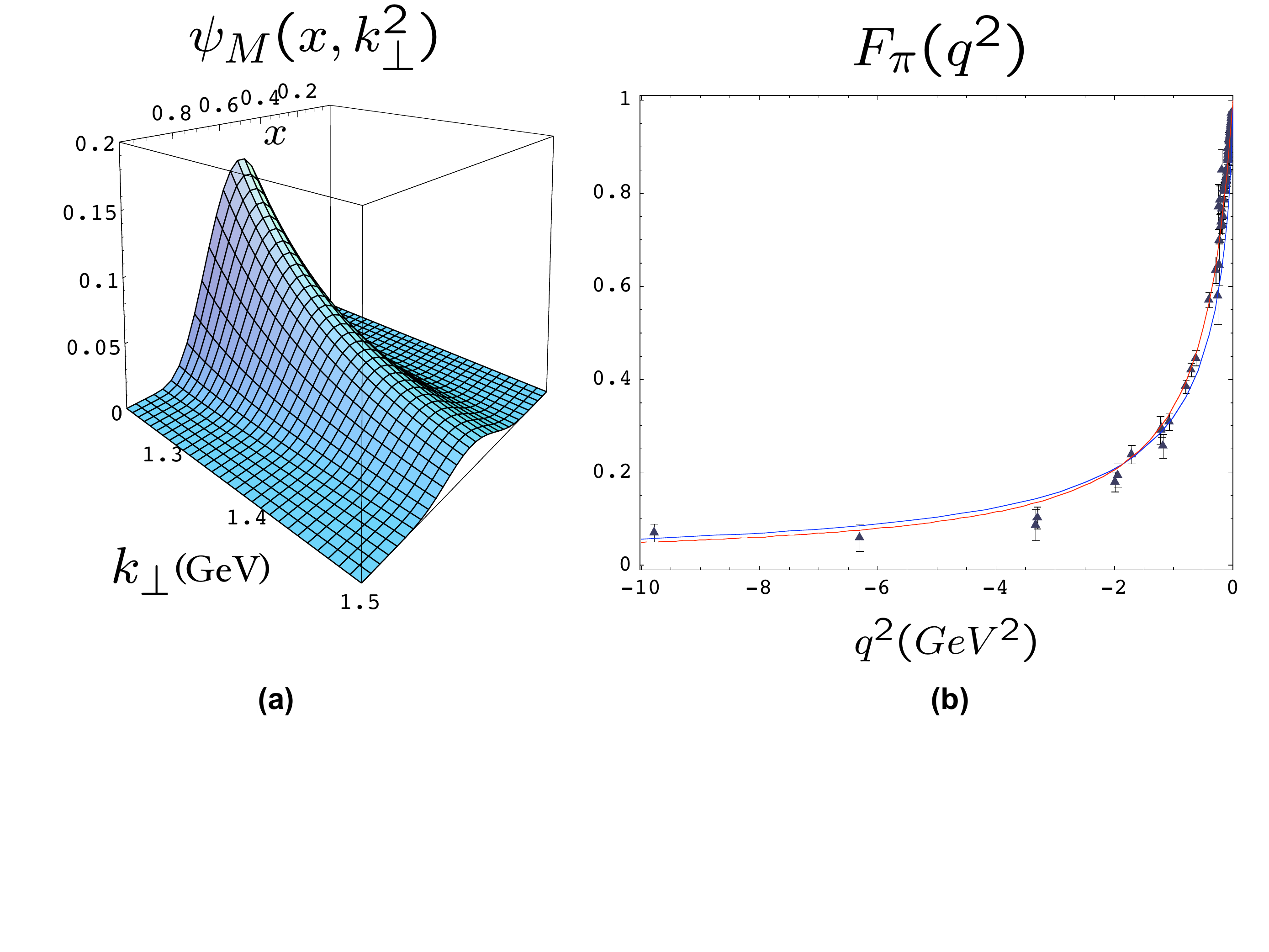}
\end{center}
\vspace{-60pt}
\caption{(a) Pion light-front wavefunction $\psi_\pi(x, \mbf{b}_\perp$)  for the  AdS/QCD  soft wall   ($\kappa = 0.375$ GeV)
model.
(b) Holographic prediction the space-like pion form factor:  (blue) hard wall ($\Lambda_{QCD} = 0.32$ GeV) and (red) soft wall  
($\kappa = 0.375$ GeV)  models.}
\label{figff}  
\end{figure}

The use of the invariant coordinate $\zeta$ in
light-front QCD allows the separation of the dynamics of quark and
gluon binding from the kinematics of constituent spin and internal
orbital angular momentum. The result is a single-variable
light-front Schr\"odinger equation  which determines the
eigenspectrum and the light-front wavefunctions of hadrons for
general spin and orbital angular momentum. This light-front wave
equation serves as a first approximation to QCD and is equivalent to
the equations of motion which describe the propagation of spin-$J$
modes on  anti-de Sitter (AdS) space. Remarkably the AdS equations
correspond to the kinetic energy terms of  the partons inside a
hadron, whereas the interaction terms build confinement and
correspond to the truncation of AdS space~\cite{deTeramond:2008ht}. As in this approximation
there are no interactions up to the confining scale, there are no
anomalous dimensions. This may explain the experimental success of
power-law scaling in hard exclusive reactions where there are no
indication of  the effects of anomalous dimensions. For the same
reason  we also expect little effect of anomalous dimensions on the
gravity side for $J > 2$. This also explains why physical hadrons
lying on Regge trajectories with $J>2$ are not incompatible with a
string description. In the hard wall model there is a total decoupling of the internal orbital angular momentum
from the total hadronic spin $J$, and thus the light-front excitation
spectrum of hadrons  depend only on the orbital and principal
quantum numbers. In the hard-wall holographic model the dependence
is linear:  $\mathcal{M}_n \sim 2n + L$. In the soft-wall model the
usual Regge behavior is found $\mathcal{M}^2 \sim n + L$. One can
systematically improve the AdS/QCD approximation by diagonalizing
the QCD LF Hamiltonian on the AdS/QCD basis or by generalizing the
variational and other systematic methods used in chemistry and
nuclear physics. The action of the non-diagonal terms in the QCD
interaction Hamiltonian generates the form of the higher Fock state
structure of hadronic LFWFs. We emphasize, that in contrast with the
original AdS/CFT correspondence, the large $N_C$ limit is not
required to connect light-front QCD to an effective dual gravity
approximation.

\section{Hadronization at the Amplitude Level}

The conversion of quark and gluon partons is usually discussed in terms  of on-shell hard-scattering cross sections convoluted with {\it ad hoc} probability distributions. 
The LF Hamiltonian formulation of quantum field theory provides a natural formalism to compute 
hadronization at the amplitude level.  In this case one uses light-front time-ordered perturbation theory for the QCD light-front Hamiltonian to generate the off-shell  quark and gluon T-matrix helicity amplitude  using the LF generalization of the Lippmann-Schwinger formalism:
\begin{equation}
T ^{LF}= 
{H^{LF}_I } + 
{H^{LF}_I }{1 \over {\cal M}^2_{\rm Initial} - {\cal M}^2_{\rm intermediate} + i \epsilon} {H^{LF}_I }  
+ \cdots 
\end{equation}
Here   ${\cal M}^2_{\rm intermediate}  = \sum^N_{i=1} {(\mbf{k}^2_{\perp i} + m^2_i )/x_i}$ is the invariant mass squared of the intermediate state and ${H^{LF}_I }$ is the set of interactions of the QCD LF Hamiltonian in the ghost-free light-cone gauge~\cite{Brodsky:1997de}.
The $T^{LF}$-matrix element is
evaluated between the out and in eigenstates of $H^{QCD}_{LF}$.   The event amplitude generator is illustrated for $e^+ e^- \to \gamma^* \to X$ in fig. \ref{fig1}.
The LFWFS of AdS/QCD can be used as the interpolating amplitudes between the off-shell quark and gluons and the bound-state hadrons.
Specifically,
if at any stage a set of  color-singlet partons has  light-front kinetic energy 
$\sum_i {\mbf{k}^2_{\perp i}/ x_i} < \Lambda^2_{QCD}$, then one coalesces the virtual partons into a hadron state using the AdS/QCD LFWFs.   This provides a specific scheme for determining the factorization scale which  matches perturbative and nonperturbative physics.

\begin{figure}[!]
 \begin{center}
\includegraphics[width=8.0cm]{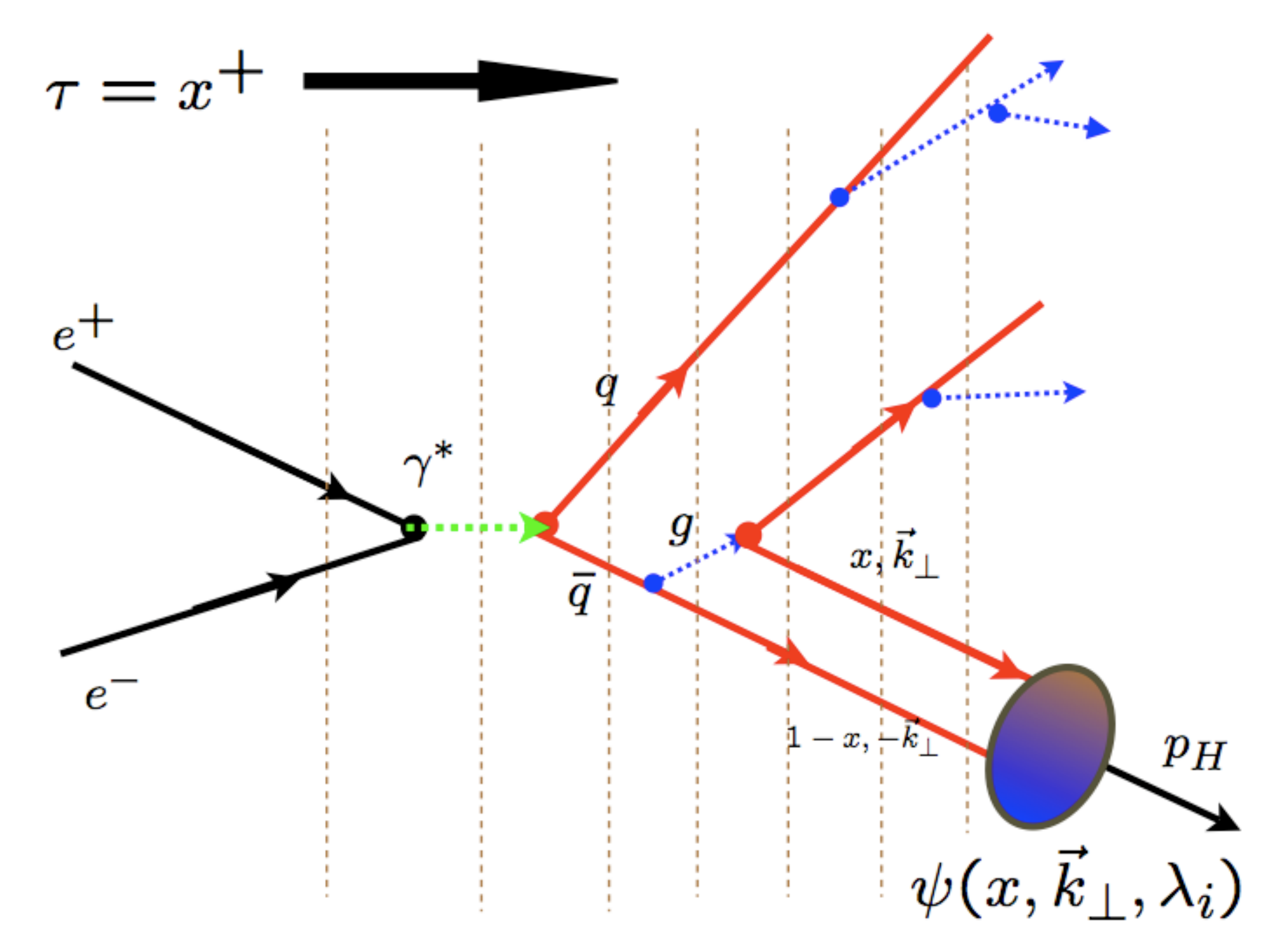}
\end{center}
\vspace{-20pt}
  \caption{Illustration of an event amplitude generator for $e^+ e^- \to \gamma^* \to X$ for 
  hadronization processes at the amplitude level. Capture occurs if
  $\zeta^2 = x(1-x) \mbf{b}_\perp^2 > 1/ \Lambda_{\rm QCD}^2$
   in the AdS/QCD hard wall model of confinement;  i.e. if
  $\mathcal{M}^2 = \frac{\mbf{k}_\perp^2}{x(1-x)} < \Lambda_{\rm QCD}^2$.}
\label{fig1}  
\end{figure}

This scheme has a number of  important computational advantages:
(a) Since propagation in LF Hamiltonian theory only proceeds as $\tau$ increases, all particles  propagate as forward-moving partons with $k^+_i \ge 0$.  There are thus relatively few contributing
 $\tau-$ordered diagrams.
(b) The computer implementation can be highly efficient: an amplitude of order $g^n$ for a given process only needs to be computed once.
(c) Each amplitude can be renormalized using the ``alternate denominator''  counterterm method~\cite{Brodsky:1973kb}, rendering all amplitudes UV finite.
(d) The renormalization scale in a given renormalization scheme  can be determined for each skeleton graph even if there are multiple physical scales.
(e) The $T^{LF}$-matrix computation allows for the effects of initial and final state interactions of the active and spectator partons. This allows for leading-twist phenomena such as diffractive DIS, the Sivers spin asymmetry and the breakdown of the PQCD Lam-Tung relation in Drell-Yan processes.
(f)  ERBL and DGLAP evolution are naturally incorporated, including the quenching of  DGLAP evolution  at large $x_i$ where the partons are far off-shell.
(g) Color confinement can be incorporated at every stage by limiting the maximum wavelength of the propagating quark and 
gluons~\cite{Brodsky:2008be}.

\section{Conclusions}

We have identified an invariant  light-front coordinate $\zeta$ which
allows the separation of the dynamics of quark and gluon binding
from the kinematics of constituent spin and internal orbital angular
momentum. The result is a single-variable light-front Schr\"odinger
equation for QCD which determines the eigenspectrum and the
light-front wavefunctions of hadrons for general spin and orbital
angular momentum. This frame-independent light-front wave equation is equivalent to
the equations of motion which describe the propagation of spin-$J$
modes on  anti-de Sitter (AdS) space~\cite{deTeramond:2008ht}.
Light-Front Holography is one of the most remarkable features of AdS/CFT.  It  allows one to project the functional dependence of the wavefunction $\Phi(z)$ computed  in the  AdS fifth dimension to the  hadronic frame-independent light-front wavefunction $\psi(x_i, \mbf{b}_{\perp i})$ in $3+1$ physical space-time. The 
variable $z $ maps  to $ \zeta(x_i, \mbf{b}_{\perp i})$. To prove this, we have shown that there exists a correspondence between the matrix elements of the energy-momentum tensor of the fundamental hadronic constituents in QCD with the transition amplitudes describing the interaction of string modes in anti-de Sitter space with an external graviton field which propagates in the AdS 
interior~\cite{Brodsky:2008pf}. The agreement of the results for both electromagnetic and gravitational hadronic transition amplitudes provides an important consistency test and verification of holographic mapping from AdS to physical observables defined on the light-front. 
The transverse coordinate $\zeta$ is related to the invariant mass squared  of the constituents in the LFWF  and its off-shellness  in  the light-front kinetic energy,  and it is thus the natural variable to characterize the hadronic wavefunction.  

It is interesting to note that the form of the nonperturbative pion distribution amplitude $ \phi_\pi(x)$ obtained from integrating the $ q \bar q$ valence LFWF $\psi(x, \mbf{k}_\perp)$  over $\mbf{k}_\perp$,
has a quite different $x$-behavior than the
asymptotic distribution amplitude predicted from the PQCD
evolution~\cite{Lepage:1979zb} of the pion distribution amplitude.
The AdS prediction
$ \phi_\pi(x)  = \sqrt{3}  f_\pi \sqrt{x(1-x)}$ has a broader distribution than expected from solving the ERBL evolution equation in perturbative QCD.
This observation appears to be consistent with the results of the Fermilab diffractive dijet 
experiment~\cite{Aitala:2000hb}, the moments obtained from lattice QCD~\cite{Brodsky:2008pg} and pion form factor data~\cite{Choi:2006ha}.

Nonzero quark masses are naturally incorporated into the AdS predictions~\cite{Brodsky:2008pg} by including them explicitly in the LF kinetic energy  $\sum_i ( {\mbf{k}^2_{\perp i} + m_i^2})/{x_i}$. Given the nonpertubative LFWFs one can predict many interesting phenomenological quantities such as heavy quark decays, generalized parton distributions and parton structure functions.  
The AdS/QCD model is semiclassical and thus only predicts the lowest valence Fock state structure of the hadron LFWF.  In principle, the model can be systematically improved by diagonalizing the full QCD light-front Hamiltonian on the AdS/QCD basis.

Color confinement and its implementation in  AdS/QCD  implies a maximal
wavelength for confined quarks and gluons and thus a finite IR fixed point for
the QCD coupling~\cite{Brodsky:2008be}. This strengthens our understanding of the narrow widths of
the $J/\psi$ and $\Upsilon$.   
A new perspective on the nature of quark and gluon condensates in
quantum chromodynamics  is presented in~\cite{Brodsky:2008xm}: the spatial support of QCD condensates
is restricted to the interior of hadrons, since they arise due to the
interactions of confined quarks and gluons.  Chiral symmetry is thus broken in a limited domain of size $1/ m_\pi$,  in analogy to the limited physical extent of superconductor phases.
This picture explains recent results which find no significant signal for the vacuum gluon
condensate.  

We also note the importance of distinguishing between static observables such as the probability distributions  computed from the square of the light-front wavefunctions versus dynamical observables such as the structure functions and the leading twist single-spin asymmetries  measured in deep inelastic scattering which include the effects of final state interactions. This distinction is summarized in fig. \ref{figstatdyn}.

\begin{figure}[!]
 \begin{center}
\includegraphics[width=12cm]{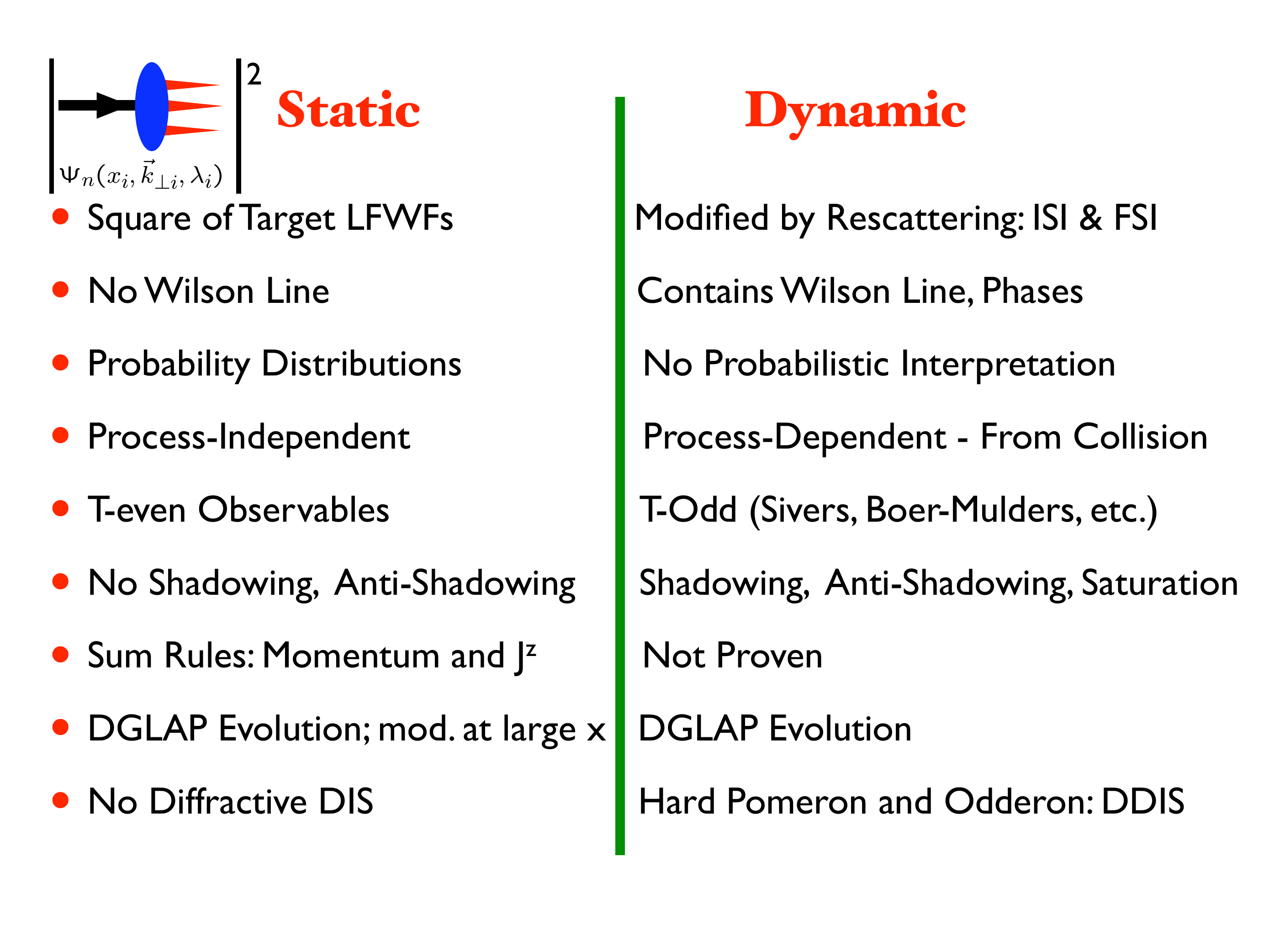}
\end{center}
\vspace{-30pt}
\caption{Dynamic versus static observables.}
\label{figstatdyn}  
\end{figure}

\section*{Acknowledgments}
Presented by SJB at Light Cone 2008: Relativistic Nuclear And Particle Physics (LC2008)
7-11 Jul 2008, Mulhouse, France.  He thanks the Institute for Particle Physics Phenomenology, Durham, UK for its hospitality.  We also thank Robert Shrock and James Vary for helpful discussions.
This research was supported by the Department
of Energy contract DE--AC02--76SF00515. SLAC--PUB--13428.


\begin{thebibliography}{99}

\bibitem{Dirac:1949cp}
  P.~A.~M.~Dirac,
  Rev.\ Mod.\ Phys.\  {\bf 21}, 392 (1949).

\bibitem{Srivastava:2002mw}
  P.~P.~Srivastava and S.~J.~Brodsky,
  Phys.\ Rev.\  D {\bf 66}, 045019 (2002)
  [arXiv:hep-ph/0202141].

\bibitem{Brodsky:2000xy}
  S.~J.~Brodsky, M.~Diehl and D.~S.~Hwang,
  Nucl.\ Phys.\  B {\bf 596}, 99 (2001)
  [arXiv:hep-ph/0009254].

\bibitem{deTeramond:2008ht}
  G.~F.~de Teramond and S.~J.~Brodsky,
  arXiv:0809.4899 [hep-ph].

\bibitem{Brodsky:2006uqa}
  S.~J.~Brodsky and G.~F.~de Teramond,
  Phys.\ Rev.\ Lett.\  {\bf 96}, 201601 (2006)
  [arXiv:hep-ph/0602252].
  
   \bibitem{Brodsky:2007hb}
  S.~J.~Brodsky and G.~F.~de Teramond,
  Phys.\ Rev.\  D {\bf 77}, 056007 (2008)
  [arXiv:0707.3859 [hep-ph]].

\bibitem{Brodsky:2008pg}
  S.~J.~Brodsky and G.~F.~de Teramond,
  arXiv:0802.0514 [hep-ph].

\bibitem{Chodos:1974je}
  A.~Chodos, R.~L.~Jaffe, K.~Johnson, C.~B.~Thorn and V.~F.~Weisskopf,
  Phys.\ Rev.\  D {\bf 9}, 3471 (1974).

\bibitem{Polchinski:2001tt}
  J.~Polchinski and M.~J.~Strassler,
  Phys.\ Rev.\ Lett.\  {\bf 88}, 031601 (2002)
  [arXiv:hep-th/0109174].

\bibitem{Karch:2006pv}
  A.~Karch, E.~Katz, D.~T.~Son and M.~A.~Stephanov,
  Phys.\ Rev.\  D {\bf 74}, 015005 (2006)
  [arXiv:hep-ph/0602229].

\bibitem{Maldacena:1997re}
  J.~M.~Maldacena,
  Adv.\ Theor.\ Math.\ Phys.\  {\bf 2}, 231 (1998)
  [Int.\ J.\ Theor.\ Phys.\  {\bf 38}, 1113 (1999)]
  [arXiv:hep-th/9711200].

\bibitem{Deur:2008rf}
  A.~Deur, V.~Burkert, J.~P.~Chen and W.~Korsch,
  Phys.\ Lett.\  B {\bf 665}, 349 (2008)
  [arXiv:0803.4119 [hep-ph]].

\bibitem{Brodsky:2008be}
  S.~J.~Brodsky and R.~Shrock,
  Phys.\ Lett.\  B {\bf 666}, 95 (2008)
  [arXiv:0806.1535 [hep-th]].
  
   \bibitem{Brodsky:2008pf}
  S.~J.~Brodsky and G.~F.~de Teramond,
  Phys.\ Rev.\  D {\bf 78}, 025032 (2008)
  [arXiv:0804.0452 [hep-ph]].

\bibitem{Brodsky:1997de}
  S.~J.~Brodsky, H.~C.~Pauli and S.~S.~Pinsky,
  Phys.\ Rept.\  {\bf 301}, 299 (1998)
  [arXiv:hep-ph/9705477].

\bibitem{Brodsky:1973kb}
  S.~J.~Brodsky, R.~Roskies and R.~Suaya,
  Phys.\ Rev.\  D {\bf 8}, 4574 (1973).

\bibitem{Lepage:1979zb}
  G.~P.~Lepage and S.~J.~Brodsky,
  Phys.\ Lett.\  B {\bf 87}, 359 (1979).

\bibitem{Aitala:2000hb}
  E.~M.~Aitala {\it et al.}  [E791 Collaboration],
  Phys.\ Rev.\ Lett.\  {\bf 86}, 4768 (2001)
  [arXiv:hep-ex/0010043].

\bibitem{Choi:2006ha}
  H.~M.~Choi and C.~R.~Ji,
  Phys.\ Rev.\  D {\bf 74}, 093010 (2006)
  [arXiv:hep-ph/0608148].

\bibitem{Brodsky:2008xm}
  S.~J.~Brodsky and R.~Shrock,
  arXiv:0803.2541 [hep-th].
  
  
  \end{thebibliography}
\end{document}